\documentstyle[preprint,aps,epsfig]{revtex}
\begin{document} 
\preprint{Nuclear Physics A (2001) in press.}
\title{Probing Mechanical and Chemical Instabilities in Neutron-Rich Matter} 
\bigskip 
\author{\bf Bao-An Li\footnote{email: Bali@astate.edu}, Andrew T. Sustich, Matt Tilley and Bin Zhang} 
\address{Department of Chemistry and Physics\\
P.O. Box 419, Arkansas State University\\
State University, Arkansas 72467-0419, USA}
\maketitle

\maketitle 
\begin{abstract}
The isospin-dependence of mechanical and chemical instabilities is  
investigated within a thermal and nuclear transport model using 
a Skyrme-type phenomenological equation of state for neutron-rich matter. 
Respective roles of the nuclear mean field and the 2-body stochastic 
scattering on the evolution of density and isospin fluctuations in either 
mechanically or chemically unstable regions of neutron-rich matter are investigated. 
It is found that the mean field dominates overwhelmingly the fast growth of both 
fluctuations, while the 2-body scattering influences significantly the later growth of the 
isospin fluctuation only. The magnitude of both fluctuations decreases with the increasing 
isospin asymmetry because of the larger reduction of the attractive isoscalar mean 
field by the stronger repuslive neutron symmetry potential in the more neutron-rich matter.
Moreover, it is shown that the isospin fractionation happens later, but grows faster 
in the more neutron-rich matter. Implications of these results to current 
experiments exploring properties of neutron-rich matter are discussed.
\end{abstract}  
\newpage 
\section{Introduction}
Explosion mechanisms of supernova and related properties of neutron stars are 
among the most interesting topics of modern nuclear astrophysics.
Investigations into these questions rely critically on the knowledge about the 
equation of state of isospin asymmetric nuclear matter. 
Thus, one of the most important subjects in nuclear physics is the 
study of the isospin-dependence of the nuclear equation of state 
using nuclear reactions induced by neutron-rich nuclei or radioactive 
beams\cite{rib,ria,li01}. In these reactions transient states of nuclear 
matter with sufficiently high isospin asymmetries as well as large thermal and 
compressional excitations can be created. The rapid progress in recent 
experiments with rare isotopes has therefore made the study of novel properties of 
extremely isospin-asymmetric nuclear matter possible. The planned Rare Isotope 
Accelerator ({\rm RIA}) will further enhance the exploration of this new frontier 
dramatically\cite{ria}. 

Prospects for discovering new physics in neutron-rich matter have generated 
much interest in the nuclear science 
community\cite{li01,li97,li98,dasgupta,tsang01,gary,sherry}. In particular, 
the isospin-dependence of the nuclear equation of state ({\rm EOS}), 
reflected mainly in the density dependence of the symmetry energy,
has recently received much attention. This is because the isospin-dependence 
of the nuclear {\rm EOS} is among the most 
important but very poorly known properties of neutron-rich 
matter\cite{wir88,brown00,hor00,dito01,lom01,bom01}. 
It is very important to the mechanisms of Type II supernova explosions and 
neutron-star mergers. It also determines the proton fraction 
and electron chemical potential in neutron stars at $\beta$ equilibrium. These 
quantities consequently influence the cooling rate of protoneutron stars and 
the possibility of kaon condensation in dense stellar 
matter\cite{lat91,bom94,sum94,lee96,pra97}. Moreover, the isospin-dependence of 
the nuclear {\rm EOS} also determines the stability boundaries of asymmetric
nuclear matter. These boundaries and their isospin-dependence subsequently
affects the multifragmentation mechanism in nuclear reactions at intermediate 
energies. Multifragmentation is thought to occur in symmetric nuclear matter 
due to the growth of instabilities triggered by density fluctuations, such as 
the Coulomb, surface and volumetric instabilities\cite{moretto}. While in 
isospin-asymmetric nuclear matter, it is interesting to study not only the
isospin-dependence of the above instabilities, but also new mechanisms for nuclear
multifragmentation, such as the chemical instability.

In this work we shall explore the isospin-dependence of both mechanical and chemical 
instabilities using a Skyrme-type phenomenological {\rm EOS} 
for neutron-rich matter within both a thermal model and a nuclear transport model.
The paper is organized as follows. In Section 2, we shall first outline the 
isospin-dependent nuclear {\rm EOS} that we use in this work, and then investigate 
relevant thermodynamical properties of neutron-rich matter using a thermal model. 
In Section 3, we shall establish boundaries of the mechanical and chemical 
instabilities and their isospin-dependence. Then, in Section 4 we study the 
dynamical growth of mechanical and chemical instabilities and associated phenomena within a
nuclear transport model. Finally, a summary and outlook are given in Section 5.

\section{Equation of State and Thermodynamical Properties of Neutron-Rich Matter}
To understand the thermodynamic properties of isospin-asymmetric nuclear matter, 
we must have a good knowledge of its EOS. The isospin-dependent part of the 
nuclear {\rm EOS} is critical for many unique properties of asymmetric matter. 
At present, nuclear many-body theories predict vastly different 
isospin-dependent nuclear {\rm EOS} depending on both the calculation 
techniques and the bare two-body and/or three-body interactions employed, 
see e.g., \cite{wir88,brown00,hor00}. 
Therefore, we use here a Skyrme-type phenomenological {\rm EOS} for asymmetric
nuclear matter. With proper parameterizations, it allows us to sample 
predictions by different many-body theories and to calculate
many thermodynamical properties analytically. 

Various studies (e.g., \cite{bom91,hub93}) have shown that the 
energy per nucleon $e(\rho,\delta)$ in asymmetric nuclear matter 
of density $\rho$ and isospin asymmetry parameter
\begin{equation} 
\delta\equiv (\rho_n-\rho_p)/(\rho_n+\rho_p) 
\end{equation}
can be approximated very well by a parabolic function in $\delta$.
At zero temperature, the $e(\rho,\delta)$ can be parameterized as 
\begin{equation}\label{ieos}
e(\rho,\delta)= \frac{a}{2}+\frac{b}{1+\sigma}u^{\sigma}+\frac{3}{5}e_F^0u^{2/3}
+S(\rho)\cdot\delta^2.
\end{equation}
In the above $u\equiv \rho/\rho_0$ is the reduced density, $e_F^0$ is the Fermi 
energy and $a=-123.6$ MeV, $b=70.4$ MeV and $\sigma=2$ corresponding to a stiff 
nuclear {\rm EOS} of isospin-symmetric nuclear matter. The last term is the 
symmetry energy whose density dependence is currently rather 
uncertain\cite{wir88,brown00,hor00,dito01,lom01,bom01}. We adopt
here a form of
\begin{equation}
S(\rho)=S_0(\rho_0)\cdot u^{\gamma},
\end{equation}
which was used by Heiselberg and Hjorth-Jensen in their recent 
studies of neutron stars\cite{hei00} with $S_0(\rho_0)=30$ MeV and $\gamma$ 
as a free parameter. We note here that a value of about $\gamma=0.6$ was obtained
by fitting to the result of variational many-body calculations\cite{hei00,akm97}. 
The $\gamma$ parameter is directly related to the $K_{sym}$ parameter of 
asymmetric nuclear matter via 
\begin{equation}
K_{\rm sym}\equiv 9\rho_0^2\frac{\partial^2 S(\rho)}{\partial \rho^2}|_{\rho
=\rho_0}=9S_0(\rho_0)\cdot\gamma(\gamma-1).
\end{equation}
The values of $K_{sym}$ predicted by many-body theories scatter
from about $-400$ {\rm MeV} to $+466$ {\rm MeV}(e.g. \cite{bom91}). 
Experimental values extracted from studying giant monopole 
resonances of asymmetric nuclei do not constrain the $K_{\rm sym}$ 
parameter either. The reported experimental values of $K_{sym}$ 
are between $-566\pm 1350$ MeV and $34\pm 159$ MeV\cite{shl93}.

Shown in Fig. 1 are the equations of state with $\gamma=$ 0.5 and 1
corresponding to $K_{sym}=$ -68 MeV and 0, respectively. The saturation points
of asymmetric nuclear matter with different $\delta$ are linked with the dashed lines 
to guide the eye. These two $\gamma$ parameters lead to rather different
saturation points especially for neutron-rich matter. The main features of the
equation of state obtained with $\gamma=0.5$ and $\gamma=1$ are surprisingly
similarly to those based on the Skyrme Hartree-Fock (SHF) and the Relativistic 
Mean Field (RMF) models\cite{oya98}, respectively. 

It is well known that the symmetry energy has a kinetic and a potential 
contribution
\begin{equation} S(\rho)=\frac{3}{5}e_{\rm F}^0u^{\frac{2}{3}}(2^{\frac{2}{3}}-1)
+V_2 
\end{equation} 
where $V_2$ is the potential contribution. The corresponding symmetry 
potential energy density is  
\begin{equation} 
W_{asy}=V_2\rho\delta^{2}, 
\end{equation} 
and the single-particle potential  
$v_{\rm asy}^{q}$ can be obtained from 
\begin{equation}\label{vasy} 
v_{\rm asy}^{q}=\frac{\partial W_{asy}}{\partial \rho_{q}}
=\pm(S_0u^{\gamma}-12.7u^{2/3})\delta +
(S_0(\gamma-1)u^{\gamma}+4.2u^{\frac{2}{3}})\delta^{2}
\end{equation} 
where ``+" and ``-" are for $q=neutron$ and $q=proton$, respectively. 
Shown in Fig.\ 2 are the symmetry energy and the corresponding symmetry potential
with three $\gamma$ parameters. The magnitude of repulsive (attactive) symmetry 
potentials for neutrons (protons) increases with both the density $\rho$ and 
isospin asymmetry $\delta$. The most important effect of $v_{asy}^q$ is to 
cause the migration of neutrons (protons) from relatively high (low) to 
low (high) density regions leading to the isospin fractionation (distillation)
effect, and this effect grows with increasing $\delta$. 

For asymmetric nuclear matter at a finite temperature $T$, 
the nucleon chemical potential $\mu_q$ corresponding to the {\rm EOS} 
of Eq.\ \ref{ieos} is given by\cite{jaqaman1}
\begin{equation}\label{muq} 
\mu_q=au+bu^{\sigma}+v_{\rm asy}^{q}+T\left[{\rm ln}(\frac{\lambda_T^3\rho_q}{2}) 
+\sum_{n=1}^{\infty}\frac{n+1}{n}b_n(\frac{\lambda_T^{3}\rho_q}{2})^n\right], 
\end{equation} 
where $\lambda_T=\left[2\pi\hbar^{2}/(m_q T)\right]^{1/2}$ is the thermal 
wavelength of a nucleon. The coefficients $b_n$ are obtained from mathematical 
inversion of the Fermi distribution function\cite{jaqaman1}. 
From the above chemical potentials for neutrons and protons, the global pressure 
for asymmetric nuclear matter can be obtained from the 
Gibbs-Duhem relation 
\begin{equation}\label{gibbs} 
\frac{\partial P}{\partial \rho}=\frac{\rho}{2}\left[(1+\delta) 
\frac{\partial \mu_n}{\partial \rho}+(1-\delta)
\frac{\partial \mu_p}{\partial \rho}\right]. 
\end{equation} 
We separate the result according to 
\begin{equation} 
P=P_0+P_{asy}+P_{kin}, 
\end{equation} 
where $P_0$ is the isospin-independent nuclear interaction contribution 
\begin{equation} 
P_0=\frac{1}{2}a\rho_0u^{2}+b\frac{\sigma\rho_0}{\sigma+1}u^{\sigma+1}, 
\end{equation} 
$P_{kin}$ is the kinetic contribution 
\begin{equation} 
P_{{\rm kin}}=T\rho\{1+\frac{1}{2}
\sum_{n=1}^{\infty} b_n(\frac{\lambda_T^{3}\rho}{4})^n\left[(1+\delta)^{n+1} 
+(1-\delta)^{n+1}\right]\}, 
\end{equation}  
and $P_{asy}$ is the contribution from the isospin-dependent 
nuclear interaction $V_2$ 
\begin{equation} 
P_{asy}=(S_0\gamma\rho_0u^{\gamma+1}-8.5\rho_0u^{\frac{5}{3}})\delta^{2}. 
\end{equation} 
The total pressure $P$ depends on the isospin asymmetry $\delta$ through
both the kinetic pressure $P_{kin}$ and the asymmetry pressure $P_{asy}$.
We compare the relative values of the three partial pressures in Fig.\ 3 for a
typical temperature of 5 MeV with $\gamma=0.5$ and $\delta=$0.2, 0.4 and 0.6, 
respectively. It is seen that the pressure increases with the increasing isospin 
asymmetry and $P_{asy}$ has an appreciable contribution to the total pressure.
The isothermal spinodal (ITS) line moves toward lower densities as the isospin 
asymmetry increases. Fig.\ 4 shows in more detail the interplay of temperature T,
isospin asymmetry $\delta$ and the $\gamma$ parameter. In each window, 
the isospin asymmetry $\delta$ goes from 0 to 1 (bottom to top) with an increment of 0.2. 
By increasing the isospin asymmetry the pressure increases as if the temperature
is increasing. It is also seen that the $\gamma$ parameter has a small
effect on the total pressure especially at high temperature because of the 
dominating role of the kinetic pressure. 
 
\section{Isospin-Dependence of Mechanical and Chemical Instabilities}
Based on the {\rm EOS} obtained from various microscopic theories 
and phenomenological models\cite{lat78,bar80,muller,liko97,baran98,baran00,cat01}, 
it has long been predicted that isospin-asymmetric nuclear matter under certain 
conditions can be mechanically or chemically unstable, i.e.,  
\begin{equation}\label{mech}  
\left(\frac{\partial P}{\partial \rho}\right)_{T,\delta}\leq 0~~~({\rm mechanical}),
\end{equation}
or
\begin{equation}
\left(\frac{\partial \mu_n}{\partial \delta}\right)_{P,T}\leq 0~~~({\rm chemical}). 
\end{equation} 
In these unstable regions, small fluctuations in $\rho$ or $\delta$ respectively 
are expected to grow. 

Mechanical instabilities have been well studied in the past for
isospin-symmetric systems, here we concentrate on probing its 
isospin-dependence. Results presented in Fig. 3 and 4 allow us to explore 
effects of the temperature, density and $\gamma$ parameter on not only 
the pressure itself, but also how these factors influence the mechanical 
instability. The latter happens in the region where 
the slope of the pressure with respect to density is negative, 
as defined in Eq. \ref{mech}. From results shown in Fig.\ 3 and 4, it is 
easy to relate the increase in temperature with a decreasing 
probability to observe a mechanical instability. It is seen that the 
mechanical instability region shrinks when either the 
temperature or the isospin asymmetry increases. Above a critical 
temperature $T_c$ determined by the condition
\begin{equation} 
\left(\frac{\partial P}{\partial\rho}\right)_{T,\delta}= 
\left(\frac{\partial^{2}P}{\partial\rho^{2}}\right)_{T,\delta}=0,
\end{equation} 
the pressure increases monotonically with density, and the mechanical 
instability disappears. To be more quantitatively, we show in Fig. 5, 
the critical temperature as a function of the isospin asymmetry 
$\delta$ with a $\gamma$ parameter of 0.5. It is interesting to 
note that the critical temperature decreases with increasing
isospin asymmetry. Thus one expects to see a shrinking mechanical instability 
region for increasingly more neutron-rich systems. It is important to stress here
that the critical temperature $T_c$ discussed above is only associated with the
disappearance of mechanical instabilities. For isospin-asymmetric nuclear matter, 
as we shall discuss in detail in the following, the system can still be chemically
unstable against density fluctuations and undergo fragmentation above 
the critical temperature $T_c$. 

To study the chemical instability and its isospin dependence we show in 
Fig. 6 the chemical potential isobars $\mu(T,P,\delta)$ as a function of 
$\delta$ for neutrons and protons with a  parameter $\gamma=0.5$ at a typical
temperature of 5 MeV and pressure of 0.005, 0.02, 0.06, 0.12, 0.2 and 0.8 
MeV/fm$^3$, respectively. We note that for corresponding pressures, 
the isobars for neutrons and protons begin from the same chemical potential 
at $\delta=0$ as one expects. The most interesting feature in this plot is 
that there exists an envelope of pressures ($0.005$ MeV/fm$^3$ $<p<$ 0.8 
MeV/fm$^3$) inside of which chemical instability occurs over a range of 
$\delta$ where  
\begin{equation} 
\left(\frac{\partial \mu_n}{\partial \delta}\right)\leq 0
\end{equation}
or
\begin{equation}
\left(\frac{\partial \mu_p}{\partial \delta}\right)\geq 0.
\end{equation} 
Inside this region, the system is unstable against isospin fluctuations. 
For instance, if several neutrons have migrated into a region of chemical 
instability due to some statistical or dynamical fluctuations in a reaction 
process, the isospin asymmetry parameter $\delta$ in the region will 
increase and the energy of that region will decrease because of its lower neutron 
chemical potential. To minimize the total energy of the system, 
it is then favorable for the system to have even more neutrons move to the 
chemically unstable region, thus leading to the further growth of the isospin 
fluctuation. Results at other temperatures show similar features. 
However, the boundary of the instability region shrinks as the temperatures T 
increase, i.e., for increasing T, chemical instability occurs for much 
less isospin asymmetric matter. The temperature dependence and its origin 
will be more clearly illustrated in the following. 
 
Having discussed separately the mechanical and chemical instabilities, we now 
compare their relative boundaries in the configure space of 
$(T,\rho,\delta)$. The Gibbs-Duhem relation of Eq.\ \ref{gibbs} can be 
used to find boundaries of mechanically unstable regions 
(ITS: isothermal spinodal) in the
$\rho-\delta$ plane for given temperatures. 
Since the chemical instability condition has to be evaluated at constant 
pressures, the following Maxwellian relation 
\begin{equation}\label{trans} 
\left(\frac{\partial \mu_n}{\partial \delta}\right)_{T,P}
=\left(\frac{\partial \mu_n}{\partial \delta}\right)_{T,\rho}
-\left(\frac{\partial \mu_n}{\partial \rho}\right)_{T,\delta} 
\cdot\left(\frac{\partial P}{\partial \rho}\right)^{-1}_{T,\delta} 
\cdot\left(\frac{\partial P}{\partial \delta}\right)_{T,\rho} 
\end{equation} 
is used to find boundaries of the chemically unstable 
regions (DS: diffusive spinodal).
Shown in Fig.\ 7 are the boundaries of the mechanical (thick lines) and 
chemical (think lines) instabilities in the $\rho-\delta$ plane with $\gamma=0.5$ 
at $T=5$ (lower window), 10 (middle window) and 15 MeV (upper window), 
respectively. It is seen that the diffusive spinodal enveloping the region of 
mechanical instability extends further out into 
the plane; the two regions of mechanical and chemical 
instabilities do no overlap. As the temperature increases, both instabilities
become less prominent over a more narrow range of densities at smaller isospin 
asymmetries. It is thus possible to observe phenomena 
due to the chemical instability even in reactions using stable nuclei at 
intermediate energies, which can attain local temperatures up to 20 MeV and 
isospin asymmetries up to about 0.4\cite{li01}.
The above features are independent of the parameters of the {\rm EOS} 
and are in good agreement with those based on more microscopic many-body 
theories\cite{lat78,bar80,muller}. In particular, we found that the variation of 
$\gamma$ parameter has very little effect on the instability boundaries. 
Therefore, in the following a constant of $\gamma=0.5$ is used. The results 
shown in Fig.\ 7 not only provide the motivations but also serve as a guidance
for our following transport model simulations.

\section{Evolution of Density and Isospin Fluctuations in Mechanically or Chemically
Unstable Neutron-Rich Matter}
Density and isospin fluctuations are expected to grow in the mechanically and/or
chemically unstable regions of asymmetric nuclear matter. Information about 
the respective growth rates of these fluctuations and how they depend on the 
isospin-asymmtry of nuclear matter is currently still rather rare. This 
information is vitally important for the structure and stability of both neutron 
stars and radioactive nuclei as well as mechanisms of nuclear multifragmentation 
in reactions induced by neutron-rich nuclei. In the following, we study the 
evolution of fluctuations in isospin-asymmetric nuclear matter initialized in 
the mechanically or chemically unstable regions within an isospin-dependent transport 
model IBUU\cite{li97,li98}. We note that Baran et al have used a similar approach 
\cite{baran98} in studying thermodynamical instabilities which can be identified
either as a mechanical or chemical instability\cite{baran00}.

The IBUU model uses consistently the isospin-dependent 
{\rm EOS} and the corresponding potentials as in the thermal model outlined 
in Section 2. Moreover, scatterings among neutrons and protons are fully 
isospin-dependent in terms of their total and differential cross sections 
as well as the Pauli blockings. Nucleons are initialized using a Boltzmann 
distribution function in a cubic box of length $L_{box}$ with periodic 
boundary conditions. The box is further divided into cells of 1 fm$^3$ volume
in which the average density $\rho_{cell}$ and isospin asymmetry $\delta_{cell}$
are evaluated. We used $10^{4}$ test particles per nucleon in evaluating the
$\rho_{cell}$ and $\delta_{cell}$. 

Shown in Fig. 8 is an illustration of the evolution of a system initialized 
in the mechanically unstable region with 
$T_i=5$ MeV, $\delta_i=0.6$ and $\rho_i=0.05$ fm$^{-3}$. For the purpose of 
this illustration a small box of $L_{box}=$10 fm is used.  
The most interesting feature shown here is the gradually increasing isospin 
fractionation. This is indicated by the spreading of the initial system into 
regions with $\rho\leq \rho_i$ and $\delta\geq\delta_i$ and where 
$\rho\geq \rho_i$ but $\delta\leq\delta_i$. The degree of isospin fractionation 
can be quantified by using the ratio $(N/Z)_{gas}/(N/Z)_{liquid}$, 
where $(N/Z)_{gas}$ and $(N/Z)_{liquid}$ is the isospin asymmetry of the low 
$(\rho/\rho_0\leq 1/8)$ and high $(\rho/\rho_0> 1/8)$ density regions, 
respectively. Shown in Fig.\ 9 are the degree of isospin fractionation as a 
function of time for a system initialized with $T_i=$5 MeV at a density of 
$\rho_i=0.05$ $fm^{-3}$. As seen from Fig.\ 7, the system is mechanically 
unstable with an isospin asymmetry $\delta$ of 0.2 and 0.6, while it is 
chemically unstable with $\delta=0.9$. As one expects from the $\delta$ 
dependence of the symmetry potential $v_{asy}$, the isospin fractionation 
grows faster with the increasing $\delta$.
It is also interesting to note that the isospin fractionation happens 
later as the $\delta_i$ increases.

The variations of the density $\rho$ and isospin asymmetry $\delta$ with 
respect to their initial values can be quantified by using, respectively, 
\begin{equation}
\sigma_d(t)\equiv(\bar{\rho^2}-\rho_i^2)^{1/2}
\end{equation}
and 
\begin{equation}
\sigma_{\delta}(t)\equiv(\bar{\delta^2}-\delta_i^2)^{1/2},
\end{equation}
where the average is over all cells. Shown in Fig.\ 10 and Fig. 11 
are the reduced variation in 
isospin asymmetry $\sigma_{\delta}(t)-\sigma_{\delta}(0)$ and 
density $\sigma_d(t)/\rho_i$ as a function of time for a system initialized 
at $T_i=5$ MeV, $\rho_i=$0.05 fm$^{-3}$ and $\delta_i=$0.2, 0.6 and 0.9,
respectively. As a reference, results are also shown for a 
system initialized with $T_i=15$ MeV and $\delta_i=0.9$ with the dash-dot 
lines (with this initial conditions the system is both mechanically and 
chemically stable as shown in Fig.\ 7). As one expects, both the isospin 
and density fluctuations stay almost constant and there is no isospin 
fractionation at all for this system. While for a system initialized in the 
mechanically (chemically) unstable region with $\delta_i=$0.2 and 0.6 
($\delta_i=$0.9), it is interesting to see that both the isospin and 
density fluctuations grow {\it faster} with the {\it decreasing} 
isospin asymmetry $\delta_i$. This finding is consistent with that found by
Baran et al in ref. \cite{baran98}. Thus, the more 
neutron-rich system in either the mechanically or chemically unstable regions
is more stable against the growth of both density and isospin fluctuations. 
This is also responsible for the later start of the isospin fractionation in 
the more neutron-rich systems as shown in Fig.\ 9. 

Why is the more neutron-rich matter more stable against both the density and 
isospin fluctuations? In the following we try to answer not only this question, but 
also explore seeds of fluctuations and their isospin dependence, 
and investigate which aspects of nuclear dynamics are important in governing the 
growth of fluctuations. The evolution dynamics of asymmetric nuclear matter is 
governed by the isospin-dependent nuclear mean filed and stochastic nuclear scatterings. 
It is necessary to study their respective roles in generating the fluctuations. 
We start with investigating the role of nucleon-nucleon scatterings.
In our approach, first-order effects of 2-body stochastic scatterings are included 
through the collision integral of the BUU equation. Second-order effects from the 
explicitly stochastic correction to the collision integral as in the Boltzmann-Langevin 
model are negelected. Thus, t here are two main seeds of fluctuations in our approach, i.e., 
the initial numerical fluctuations from the random sampling of the initial state and the later 
2-body stochastic nuclear scatterings. 
Both of them may lead to the growth of fluctuations by propagating  
through the nuclear mean field in the early and later stages of the evolution, 
respectively. The magnitude of the initial numerical fluctuations in both density 
and isospin asymmetrty is clearly shown in the upper window of Fig. 8. The seeds of 
fluctuations due to 2-body stochastic nuclear scatterings are determined by the 
isospin-dependent collision dynamics in asymmetric nuclear matter. 
Most importantly, the experimental neutron-neutron cross section, 
as shown in Fig.\ 12, is only about $1/3$ of that for neutron-proton 
collisions\cite{nndata}, and also the final state for neutron-neutron 
scatterings is more strongly Pauli blocked in the more neutron-rich matter. 
To be more quantitative, we show in Fig.\ 13 the Pauli blocking rate 
as a function of time. As the $\delta_i$ increases from 0.2 to 0.9 the 
Pauli blocking rate increases from about 35\% to 50\%. Therefore, there are
significantly less nucleon-nucleon collisions in more neutron-rich nuclear
matter. Shown in Fig.\ 14 are the average number of 
possible (before using the Pauli blocking) nucleon-nucleon collisions 
per nucleon $N_{try}/A$ and the successful ones ($N_{coll}/A$) as a function 
of time. The value of $N_{try}/A$ reflects directly the isospin-dependent 
elementary nucleon-nucleon cross sections shown in Fig.\ 12. 
Before about 40 fm/c, there is essentially no collision 
because of the strong Pauli blocking when the phase space nonuniformity due to the 
initial fluctuation is still small. Later, nuclear scatterings become important, moreover, 
they are more frequent with the decreasing isospin asymmetry. With $\delta_i=0.2$ each nucleon 
can make about 4 scatterings up to the time of 100 fm/c, while it can only make about 1 
collision with $\delta_i=0.9$. Because of the combined effects of both the Pauli blocking and 
the isospin-dependent elementary cross sections, the number of nuclear 
scatterings in nuclear matter with $\delta_i=0.9$ is consequently only 
about 1/6 of that with $\delta_i=0.2$, contributing to the smaller isospin 
and density fluctuations in the more neutron-rich matter. 
To separate the respective roles of the nuclear mean field and the 2-body scatterings, 
we have performed studies by turnning off the nucleon-nucleon collisions in the model. 
As shown in the upper window of Fig.\ 15,  the collisional seeds of fluctuations  
lead to the significant growth of the isospin fluctuation in the later stage of the evolution.
However, they have very little effect on the growth of density fluctuations as shown in the lower window. 
A comparison of the results obtained with and without the collision integral 
indicates that the growth of fluctuations are overwhelmingly dominated by the nuclear mean field.
This is in agreement with the analysis based on the linear response theory by Baran et al\cite{baran98}. 
The observed isospin-dependence of the fluctuations can be understood from the interplay 
between the {\it attractive isoscalar} mean field and the {\it repulsive symmetry potential for neutrons}. 
The symmetry potential, as shown in Eq.\ \ref{vasy}, is repulsive for neutrons and attractive for protons 
and their magnitudes increase with the increasing isospin asymmetry. Thus, the resultant attractive mean 
field is weaker in the more neutron-rich matter. Because of the small number of scatterings, particularly 
in the early stage of the evolution, the growth of fluctuations is thus mainly determined by the strength and 
sign of the resultant nuclear mean field according to the linear response theory. Based on the latter, 
the magnitude of both fluctuations can grow larger with the increasing strength of the attractive resultant 
mean field, and thus also with the decreasing isospin asymmetry $\delta$. 

\section{Summary and Outlook}
In summary, utilizing a phenomenological nuclear equation of state within 
a thermal model, first we explored the isospin dependence of the 
chemical and mechanical instabilities in asymmetric nuclear matter. The diffusive 
spinodal is found to extend further out into the $(\delta-\rho)$ plane 
than the isothermal spinodal. It is shown that the isospin dependence 
of the nuclear equation of state plays a key role in determining the 
thermodynamical properties of asymmetric nuclear matter. We also investigated the 
evolution of density and isospin fluctuations in mechanically and 
chemically unstable asymmetric nuclear matter within a nuclear 
transport model. Both the isospin and density fluctuations are found to 
decrease with the increasing isospin asymmetry of the system. 
Respective roles of the nuclear mean field and the 2-body stochastic 
scattering on the evolution of density and isospin fluctuations in either 
mechanically or chemically unstable regions of neutron-rich matter are investigated. 
It is found that the mean field dominates overwhelmingly the fast growth of both 
fluctuations, while the 2-body scattering influences significantly the later growth of the 
isospin fluctuation only. We have also shown that the isospin fractionation happens later, 
but grows much faster in the more neutron-rich matter. 

It is well known that more neutron-rich systems are less bound and have smaller
saturation densities as illustrated by the saturation lines in Fig. 1. However, 
fluctuations and their growth are also important 
for determining the final state of a neutron-rich system, such as in the 
projectile fragmentation in producing exotic beams\cite{fri00}. Our results above 
indicate that fluctuations actually have a compensating role to the lower 
binding energy in stabilizing neutron-rich system where fluctuations  
are smaller and do not grow as fast as in symmetric ones. 
Moreover, our results on the isospin fractionation accompanying the evolution 
of fluctuations indicate that the configuration of a more 
dense, isospin-symmetric region surrounded by a more isospin-asymmetric gas 
as in halo nuclei is a natural result of the isospin-dependent nuclear dynamics. 
Furthermore, we expect that the multifragmentation and isospin fractionation 
in nuclear reactions induced by neutron-rich nuclei to happen on longer time 
scales compared to symmetric reactions of the same masses. These expectations 
can be tested by measuring products of multifragmentation. In 
particular, the measurement of neutron-neutron, proton-proton as well as 
fragment-fragment correlation functions, in comparative studies of isospin 
symmetric and asymmetric nuclear reactions will be very 
useful\cite{robert,bauer}. 

\section{Acknowledgement}
We would like to thank V. Baran, M. Colonna, M. Di Toro, A. Evans, J.B. Natowitz, M.B. Tsang 
and S.J. Yennello for useful discussions. 
This work was supported in part by the National Science Foundation Grant 
No. 0088934 and Arkansas Science and Technology Authority Grant No. 00-B-14.

\newpage 

\begin{figure}[htp] 
\centering \epsfig{file=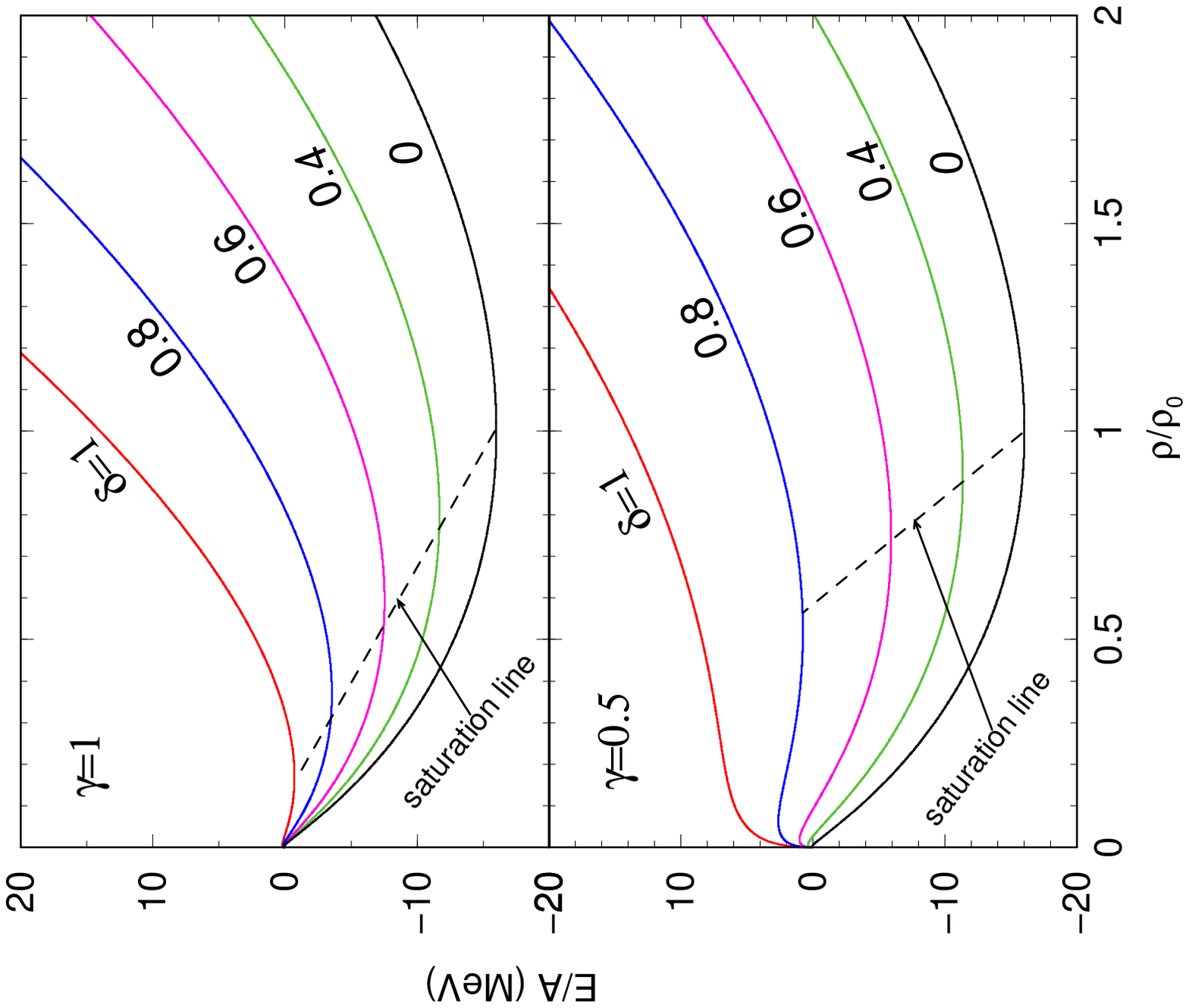,width=12cm,height=12cm,angle=-90} 
\caption{The equation of state of isospin-asymmetric nuclear matter with the
isospin asymmetry $\delta$ spans between 0 and 1 with a $\gamma$ parameter
of 0.5 (lower window) and 1 (upper window), respectively.} 
\label{fig1} 
\end{figure} 

\begin{figure}[htp] 
\centering \epsfig{file=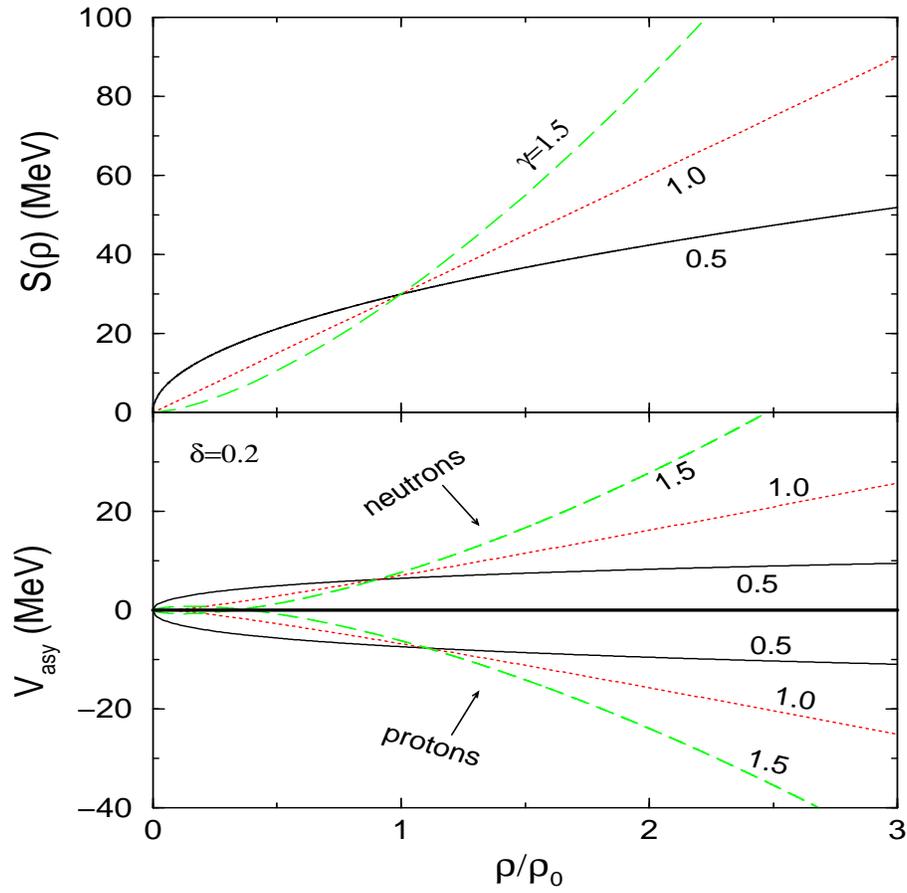,width=12cm,height=12cm,angle=-90} 
\caption{Symmetry energy (upper window) and potential (lower window) as a 
function of reduced density with a $\gamma$ parameter of 0.5, 1.0 and 1.5, 
respectively. The symmetry potential is plotted for an isospin asymmetry 
$\delta=0.2$.} 
\label{fig2} 
\end{figure} 

\begin{figure}[htp] 
\centering \epsfig{file=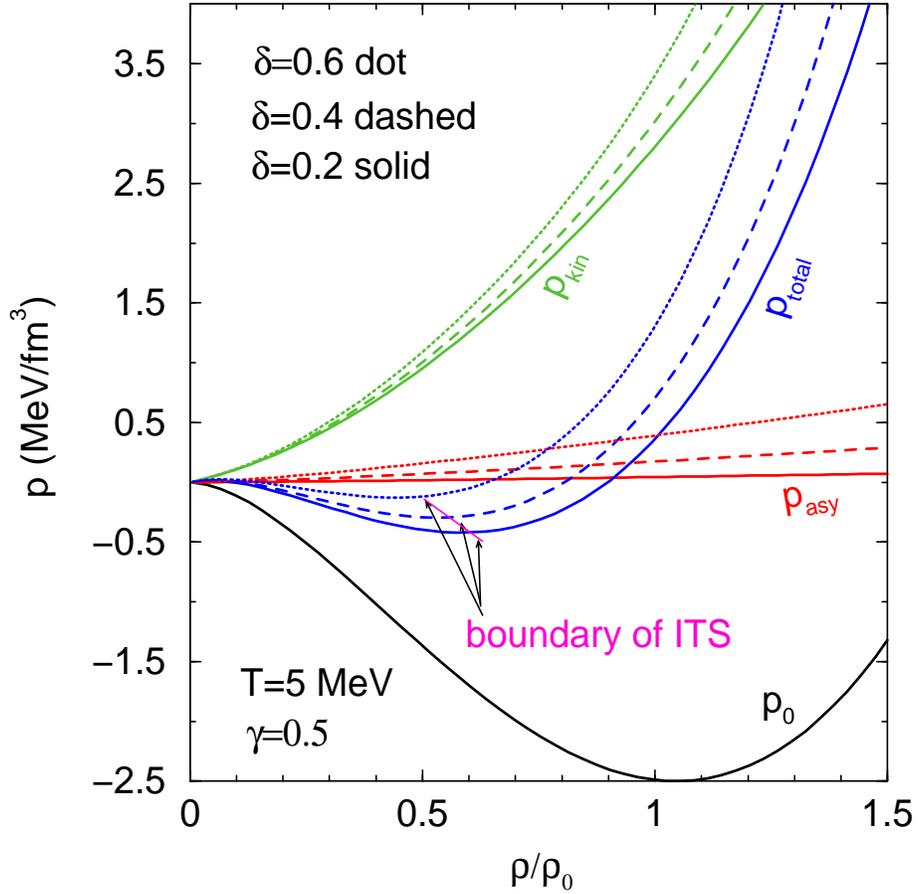,width=12cm,height=12cm,angle=-90} 
\caption{Total and partial pressures as a function of density, $\rho$, for 
isospin asymmetry $\delta=0.2,0.4$ and $0.6$ with a temperature $T=5$ MeV 
and $\gamma=0.5$. The isothermal spinodal line (ITS) is indicated with arrows.} 
\label{fig3} 
\end{figure} 

\begin{figure}[htp] 
\centering \epsfig{file=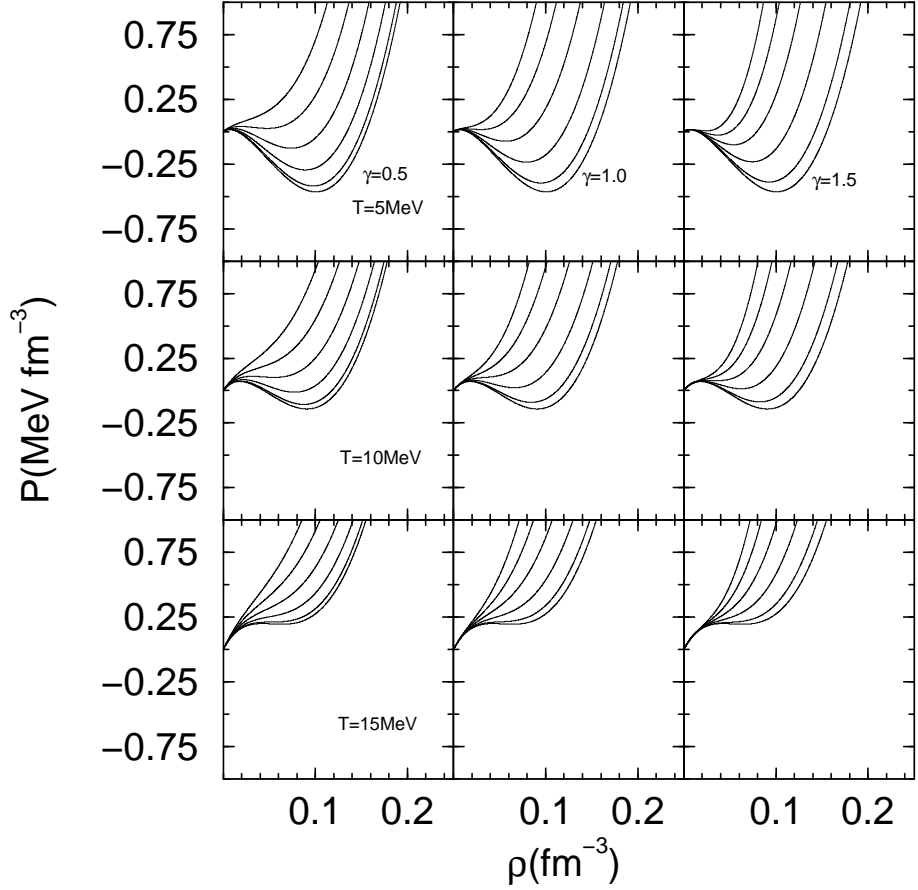,width=12cm,height=12cm,angle=-90} 
\caption{Total pressure P as a function of density $\rho$ for 
isospin asymmetry $\delta=0.0,0.2,0.4,0.6,0.8$ and $1.0$ (from bottom to top)
with a temperature $T=5, 10$ and $15$ MeV and $\gamma=0.5, 1.0$ and 1.5, 
respectively.} 
\label{fig4} 
\end{figure} 

\begin{figure}[htp] 
\centering \epsfig{file=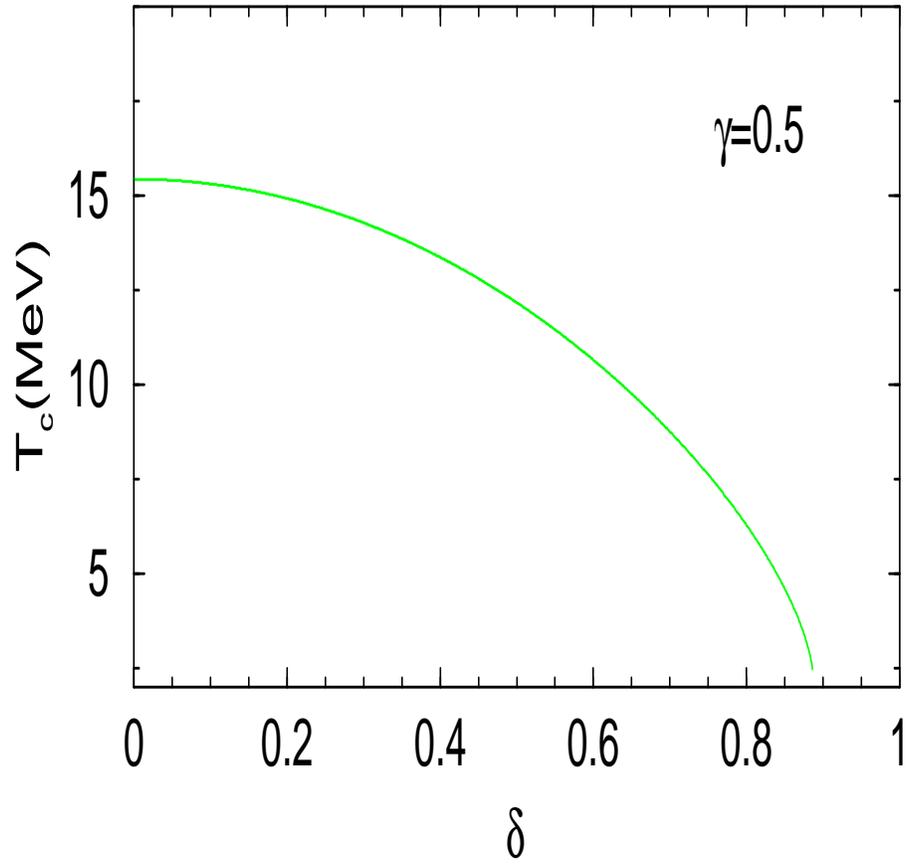,width=12cm,height=12cm,angle=-90} 
\caption{Critical temperature as a function of isospin asymmetry 
$\delta$.} 
\label{fig5} 
\end{figure}  

\begin{figure}[htp] 
\centering \epsfig{file=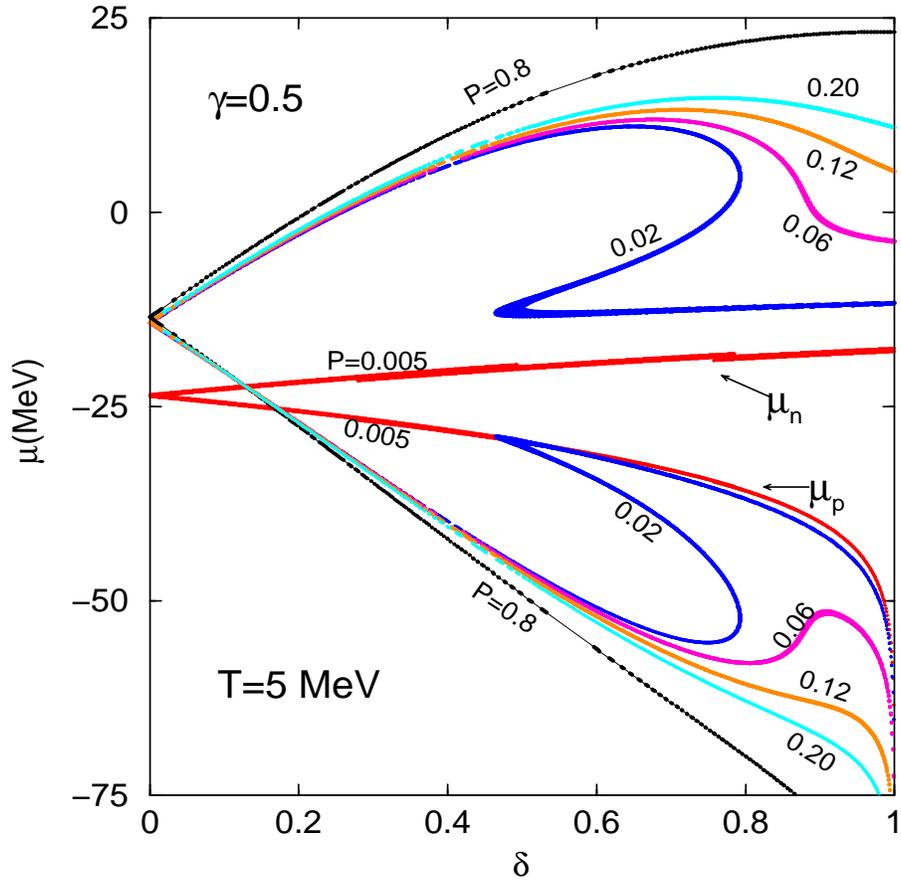,width=12cm,height=12cm,angle=-90} 
\caption{Chemical potential isobars, $\mu$ for neutrons and protons as a function 
of isospin asymmetry, $\delta$ at temperature, T=5 MeV and density parameter, 
$\gamma=0.5$. The isobars are at pressures of $P=0.005, 0.02, 0.06, 0.12, 0.20$ and $0.80$ 
MeV/fm$^3$, respectively.} 
\label{fig6}  
\end{figure} 

\begin{figure}[htp] 
\centering \epsfig{file=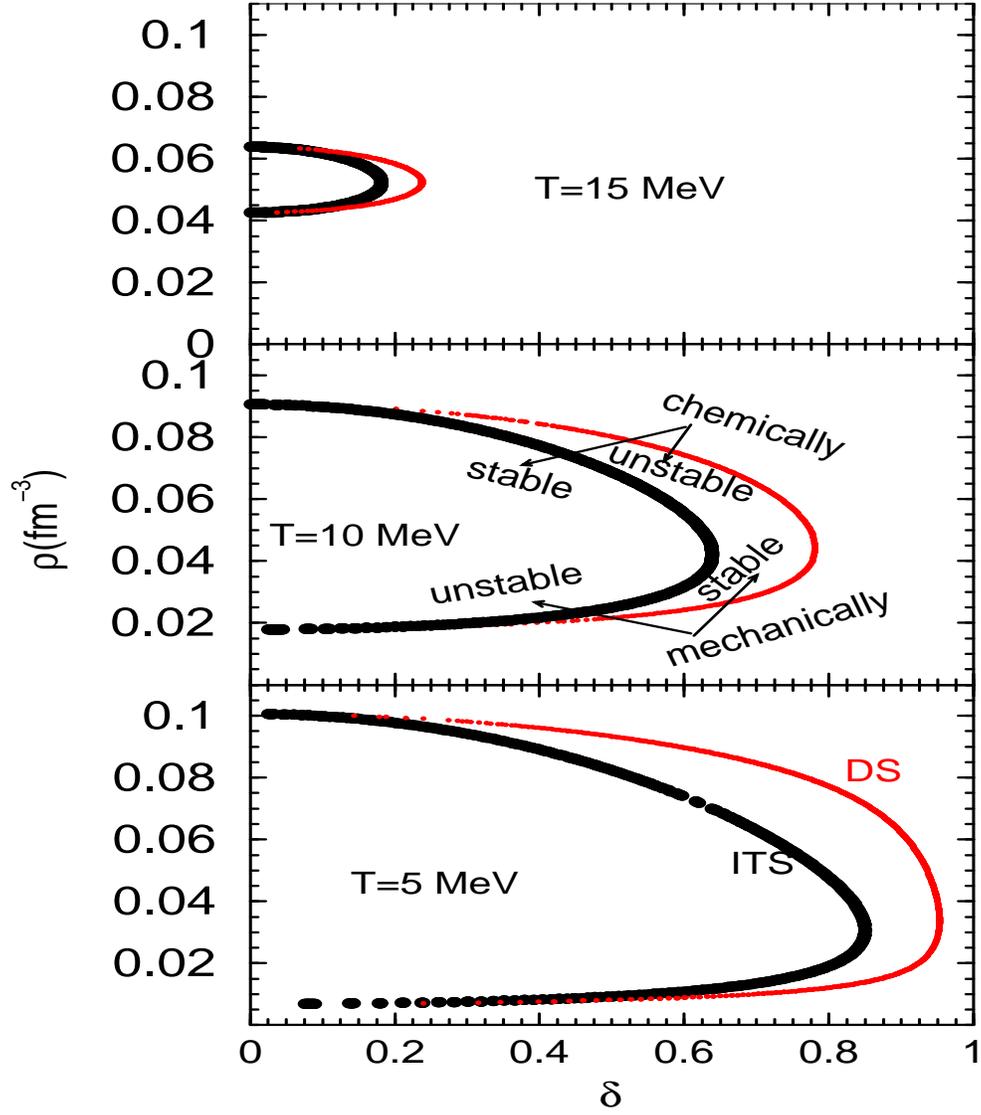,width=15cm,height=13cm,angle=-90} 
\caption{Boundaries of the mechanical (thick lines) and chemical (thin lines)
in the density-isospin asymmetry plane at a temperature of 5, 10 and 15 MeV, 
respectively. }
\label{fig7} 
\end{figure} 

\begin{figure}[htp] 
\centering \epsfig{file=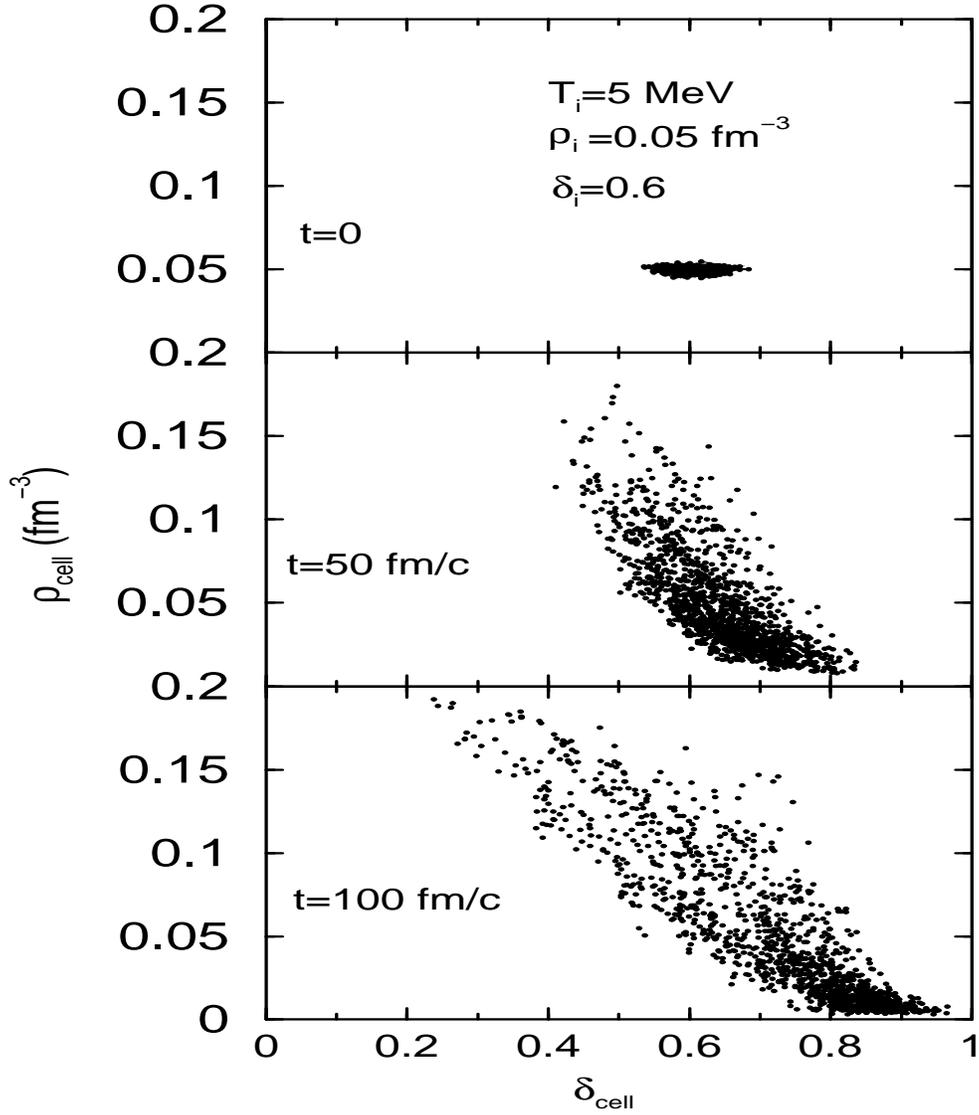,width=15cm,height=13cm,angle=-90} 
\caption{An illustration of the evolution of asymmetric nuclear matter
in the density-isospin asymmetry plane. Each point in the scatter plots 
represent one cell of 1 fm$^3$ volume in a cubic box of side length 
$L_{box}=10$ fm.}
\label{fig8}
\end{figure}   

\begin{figure}[htp] 
\centering \epsfig{file=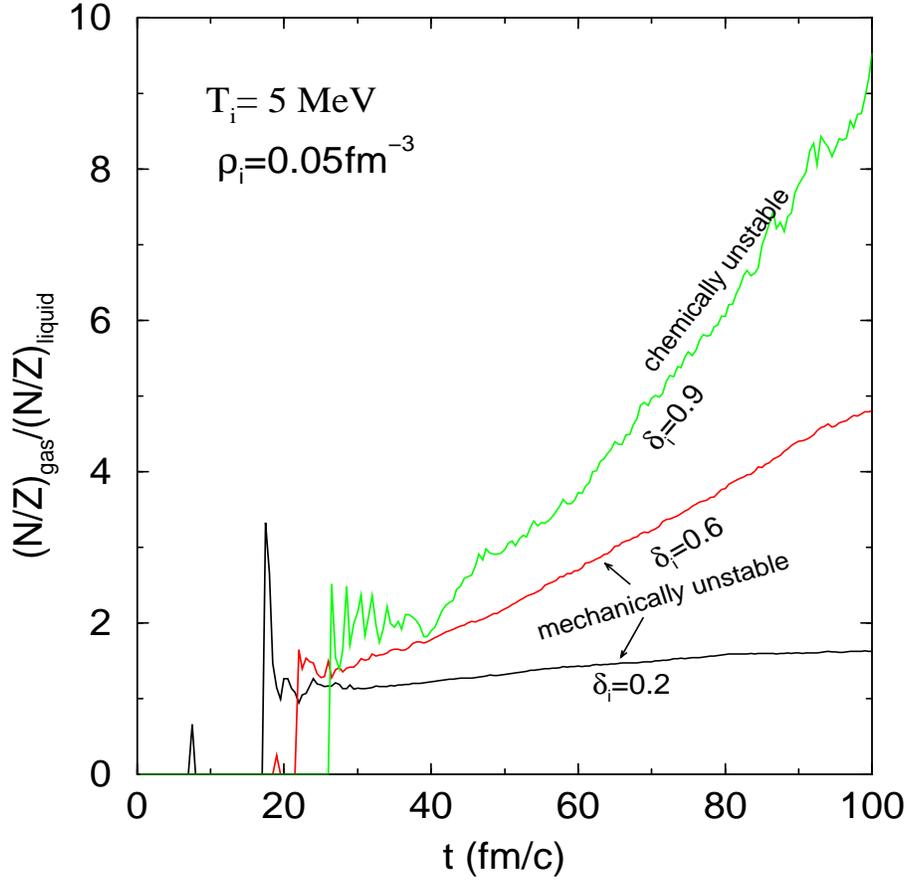,width=12cm,height=12cm,angle=-90} 
\caption{Evolution of the degree of isospin fractionation in a cubic box of side 
length $L_{box}=30$ fm and density 0.05 fm$^{-3}$, temperature $T_i=5$ MeV 
and $\delta_i=$0.2, 0.6 and 0.9, respectively.}   
\label{fig9}
\end{figure} 

\begin{figure}[htp] 
\centering \epsfig{file=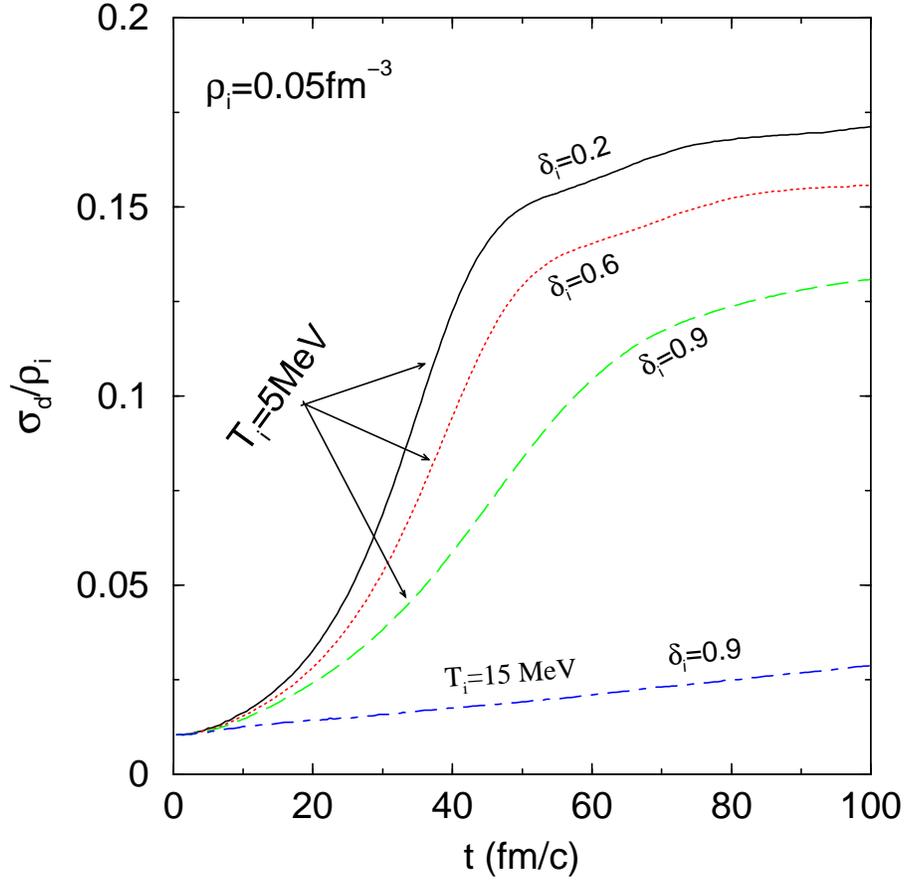,width=12cm,height=12cm,angle=-90} 
\caption{Evolution of the reduced density fluctuation in a cubic box of side 
length $L_{box}=30$ fm and density 0.05 fm$^{-3}$. The dash-dot lines 
are calculated with $T_i=15$ MeV and $\delta_i=0.9$; while the solid 
lines are calculated with $T_i=5$ MeV and $\delta_i=$0.2, 0.6 and 0.9, 
respectively.}
\label{fig10}   
\end{figure}  
 
\begin{figure}[htp] 
\centering \epsfig{file=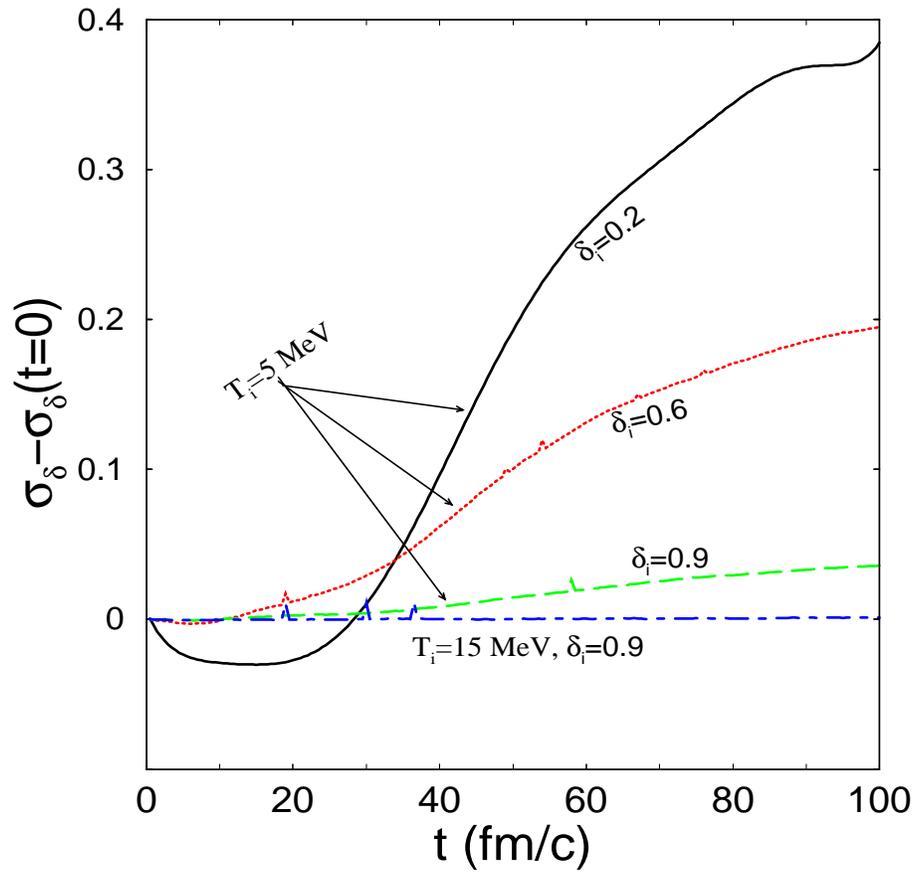,width=12cm,height=12cm,angle=-90} 
\caption{Evolution of the reduced isospin fluctuation for the same systems
as in Fig.\ 10.}
\label{fig11}   
\end{figure}  

\begin{figure}[htp] 
\centering \epsfig{file=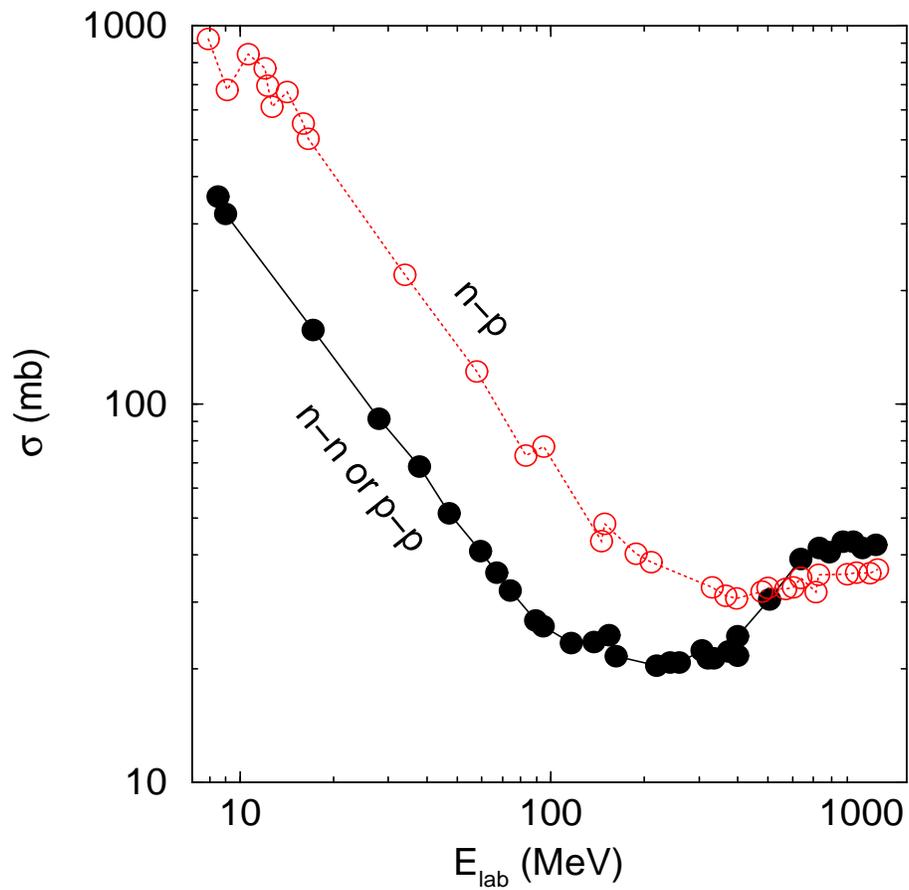,width=12cm,height=12cm,angle=-90} 
\caption{Experimental neutron-neutron and neutron-proton cross sections 
as a faction of beam energy.}
\label{fig12}
\end{figure}  

\begin{figure}[htp] 
\centering \epsfig{file=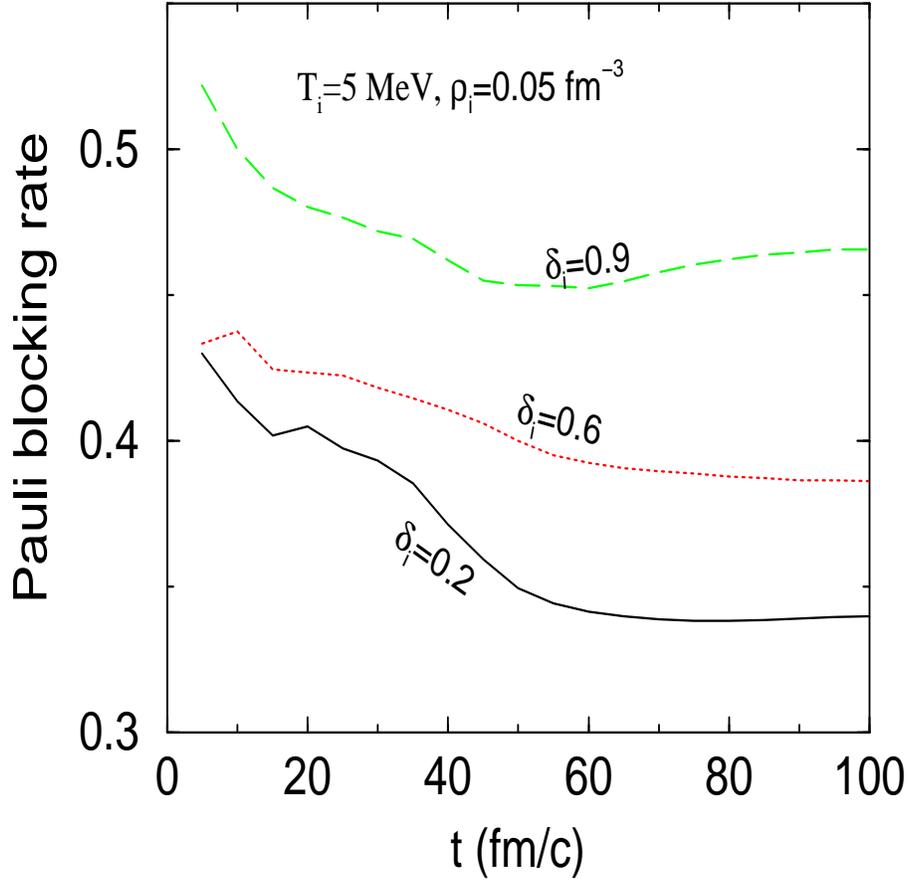,width=12cm,height=12cm,angle=-90} 
\caption{The Pauli blocking rate as a function of time in a cubic box of 
$L_{box}=30$ fm with density 0.05 fm$^{-3}$, temperature $T_i=5$ MeV and 
$\delta_i=$0.2, 0.6 and 0.9, respectively.}
\label{fig13}
\end{figure} 

\begin{figure}[htp] 
\centering \epsfig{file=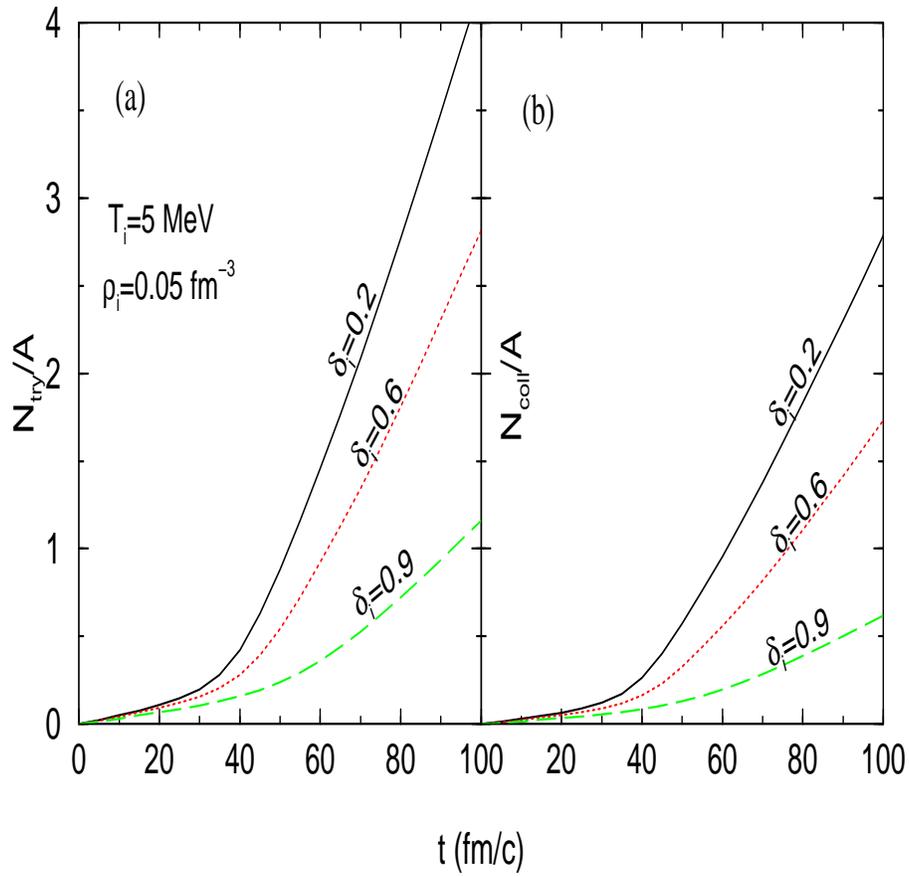,width=12cm,height=12cm,angle=-90} 
\caption{The average number of possible nucleon-nucleon collisions 
per nucleon (a) and the average number of successful nucleon-nucleon 
collisions per nucleon (b) as a function of time for the same systems as 
in Fig. 13.}
\label{fig14}
\end{figure} 

\begin{figure}[htp] 
\centering \epsfig{file=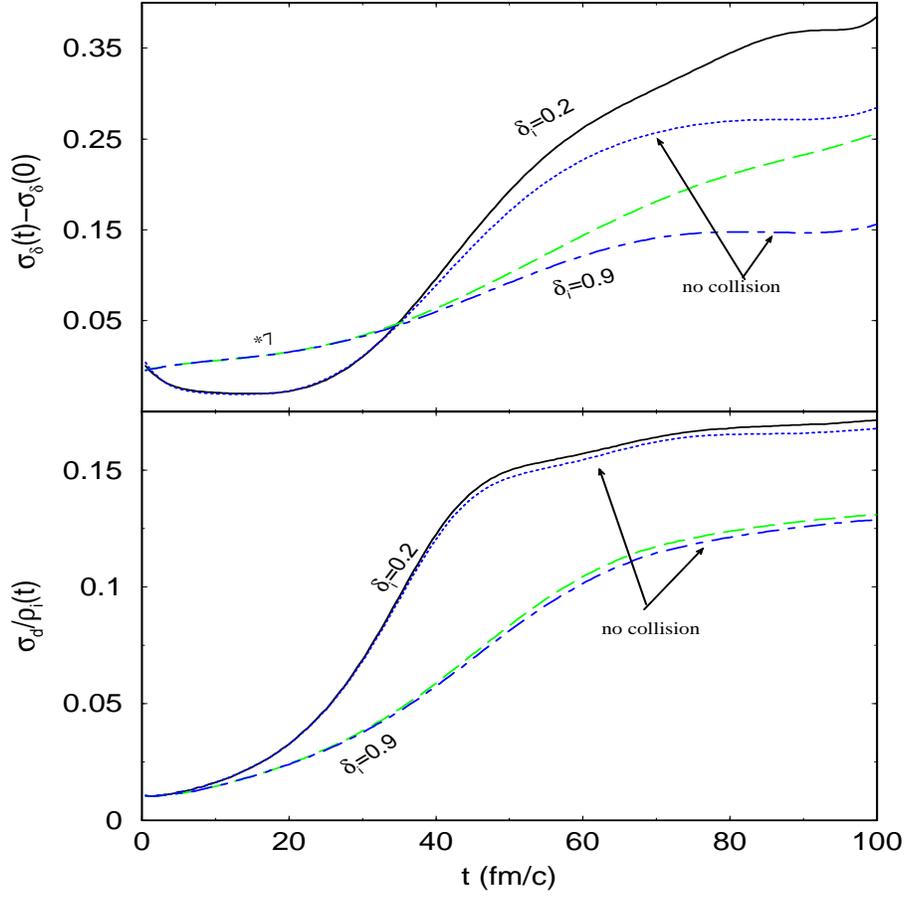,width=12cm,height=12cm,angle=-90} 
\caption{Evolution of isospin (upper window) and density (lower window)
fluctuations as a function of time. The solid and long dashed lines are
results of full calculations, while the dotted and dash-dot lines are 
results obtained by turnning off the 2-body scatterings.}
\label{fig15}
\end{figure} 

\end{document}